\documentclass[preprint2]{aastex}
\usepackage{natbib}
\usepackage{graphics}
\usepackage{subfigure}
\usepackage{graphicx}
\usepackage{amssymb}
\usepackage{hyperref}

\newcommand{\Msun}{$~\rm{M}_\odot$}

\shorttitle{SNIa in hydrodynamical simulations}
\shortauthors{Jimenez et al.}
\begin{document}

\title{Progenitors of Supernovae Type Ia  and Chemical Enrichment in
  Hydrodynamical Simulations -I. The Single Degenerate Scenario}

\author{Noelia Jim\'enez\altaffilmark{1,2,5,8},
  Patricia B. Tissera\altaffilmark{2,3,4}, and  Francesca Matteucci\altaffilmark{5,6,7}}

\email{nj22@st-andrews.ac.uk}
\altaffiltext{1}{School of Physics \& Astronomy, University of St Andrews, North Haugh,
St Andrews, KY16 9SS, Scotland, UK}
\altaffiltext{2}{Instituto de Astronom\'ia y F\'isica del
  Espacio (IAFE, CONICET-UBA), CC. 67 Suc. 28, C1428ZAA, Ciudad de Buenos Aires,
  Argentina}
\altaffiltext{3}{Departamento de Ciencias Fisicas,
  Universidad Andres Bello, Av. Republica 220,
  Santiago,Chile}
\altaffiltext{4}{Millenium Institute of Astrophysics,
  Santiago, Chile}
\altaffiltext{5}{Dipartimento di Fisica, Universita' di Trieste,
  Via G. B. Tiepolo, 11, 34100, Trieste}
\altaffiltext{6}{INAF,Trieste, Via
  G. B. Tiepolo, 11, 34100, Trieste}
\altaffiltext{7}{I.N.F.N., Via Valerio 2, 34100, Trieste}
\altaffiltext{8}{Institut d' Estudis Espacials de Catalunya (IEEC), Institut de Ciencies de L'Espai (ICE), Spain}
\begin{abstract}

The nature of the Type Ia supernovae (SNIa) progenitors remains still
uncertain. This is a major issue for galaxy evolution models since
both chemical and energetic feedback play a major role in the gas
dynamics, star formation and therefore in the overall stellar
evolution. The progenitor models for the SNIa available in the
literature propose different distributions for regulating the
explosion times of these events. These functions are known as the
Delay Time Distributions (DTDs). This work is the first one in a
series of papers aiming at studying five different DTDs for
SNIa. Here, we implement and analyse the Single Degenerate scenario
(SD) in galaxies dominated by a rapid quenching of the star formation,
displaying the majority of the stars concentrated in the bulge
component.  We find a good fit to both the present observed SNIa rates
in spheroidal dominated galaxies, and to the [O/Fe] ratios shown by
the bulge of the Milky Way.  Additionally, the SD scenario is found to
reproduce a correlation between the specific SNIa rate and the
specific star formation rate, which closely resembles the
observational trend, at variance with previous works. Our results
suggest that SNIa observations in galaxies with very low and very high
specific star formation rates can help to impose more stringent
constraints on the DTDs and therefore on SNIa progenitors.  
\end{abstract}

\keywords{galaxies: general-- galaxies: abundances -- galaxies: evolution --
  Galaxy:bulge--methods: numerical--hydrodynamics}

%\slugcomment{Submitted \today}

\section{Introduction}\label{Sec:Intro}

 Galaxy formation represents a multi-scale highly non-linear process.
 The modeling of the observed galaxy populations with a
 self-consistent model -- from the initial conditions left behind the
 Big Bang--, requires knowledge of the co-evolution of all the
 components of the Universe. According to the $\Lambda$-Cold Dark
 Matter ($\Lambda$CDM) paradigm, baryons condensate onto the dark
 matter haloes, which constitute the sites of galaxy formation. The
 theories of galaxy formation aim at understanding at the same time
 the large scale growth of the structure and small-scales processes
 such as the star formation from molecular clouds, for example. This
 is a challenging task which can be tackled by using semi-analytical
 or fully cosmological models. Both, resort to recipes or sub-grid
 modeling to include complex physical processes which cannot be
 numerically resolved. Hence, is of utmost importance to confront the
 model results with observations to learn about galaxy formation and
 to test the validity of the adopted hypotheses.

 Galaxy chemical evolution models provides us with a powerful tool to
 understand the way in which galaxies formed and evolved
 \citep{Tinsley79,MatteucciGreggio86,Matteucci94}.  Since the
 information given by chemical abundances is the result of the
 nucleosynthesis production of the stellar populations and large
 scale physics involved in the galaxy formation, chemical patterns can
 provide stringent constraints for galaxy formation models.

\begin{sloppypar}
The treatment of chemical evolution with the inclusion of different
models for Type Ia supernovae (SNeIa) opened the route to study galaxy
formation by probing links between chemical patterns and relevant
physical processes.  These efforts are usually attempted with a
variety of numerical techniques, such as multi-zone chemical evolution
models \citep[e.g.][]{MatteucciGreggio86,Isern93,Cris01,PM04},
semi-analytic models of galaxy formation
~\citep[e.g.][]{CaluraMenci09,Arrigoni10,Jimenez11,Yates13, DeLucia14,
  Tuna15}, and cosmological hydrodynamical simulations
~\citep[e.g.][to name a few]{Raiteri96,
  Carraro98,Mosconi01,Tornatore04,Nakasato03,
  Kobayashi04,SC05,Nagashima05,Wiersma09, Aumer13, Few14}.  Nowadays,
sophisticated computational tools and detailed observations of local
and high redshift galaxies enable us to improve galaxy formation
models and develop more realistic schemes for the baryonic
astrophysics.
\end{sloppypar}

One of the major astrophysical problems is the uncertain identity of
the SNIa progenitors. This is a matter of concern not only in galaxy
formation theory but also in modern cosmology. The remarkable
similarity shown by the light curves of the SNIa have made them
excellent cosmological distance indicators. Hence, they are used to
investigate the properties of dark energy, and to test parameters
of the cosmological model and the acceleration epoch of the Universe
\citep{Perlmutter99}. However, given the still unknown nature of the
progenitor systems of SNIa, systematic errors in the deduced
distances based on calibrations of nearby SNeIa, could have an impact
in the cosmological parameter estimation \citep{PanDulce12, Rest13}.

 It is well known that core-collapse supernovae (SNeII) are produced
 by massive short-lived stars (M$>8$\Msun). The nucleosynthesis
 production of these events feeds the interstellar medium (ISM) with
 energy and, mainly, with the so-called $\alpha$-elements -- O, Ne,
 Mg, Si, S and Ca --. On the other hand, the SNeIa are the main
 contributor of the Fe in the Universe \citep{Greggio83,
   MatteucciGreggio86, Cappellaro97}. This element is usually
 considered to be a tracer of the metallicity in stars.  Furthermore,
 the production and distribution of the chemical elements in the ISM
 affects the cooling rates of the gas \citep{ShuterlandDopita93}, the
 star formation processes, the subsequent stellar evolution
 \citep{Pietrinferni06}, and the production of high-redshift dust
 \citep{Maiolino04,Bianchi07}. Main properties such as the luminosity
 function and the mass distribution of galaxies might consequently be
 affected \citep{WhiteFrenk91,SC05}. By studying the enhancement of
 the $\alpha$-elements relative to Fe, we expect to learn about the
 Initial Mass Function (IMF) of the stellar populations and, very
 importantly, about the time-scales in which SNeIa become relevant.
 The time-delay between core-collapse SNe and SNeIa injection of
 chemical elements in the ISM, creates important patterns which can be
 used as clocks to tag particular events in the formation histories of
 galaxies.  Given a starburst, SNIa events occur following a
 distribution of explosion times which is known as the Delay Time
 Distribution (DTD).  Different scenarios for SNIa progenitors produce
 different DTDs which, in turn, create particular chemical patterns in
 the stellar populations and their host galaxies.

 It has long been known \citep{HoyleFowler60} that the explosion
 mechanisms, the chemical and energetic composition of the remnants
 and the light curves of SNeIa (with absence of He and H) involve, at
 the most fundamental level, the combustion of a degenerate stellar
 core. Specifically, a White Dwarf (WD) star of Carbon (C) and Oxygen
 (O) is lead away from the equilibrium, following mass accretion and
 successive explosion.  Among the two most popular scenarios for
 explaining SNIa based on a thermonuclear explosion of a C-O WD, there
 is the Single Degenerate (SD) scenario -- a WD exceeding the
 Chandrasekhar mass through accretion from a non-degenerate companion
 star-- where the mass accretion can assume many configurations. The
 secondary star is proposed to be a Main Sequence star, a subgiant
 star, a helium star or red giant star
 \citep{WhelanIben73,IbenTutukov84,VandenHeu92,Hachazo96,Hachazo99,Hansolo04,Geier13}.
 In addition to the variety of possible secondary stars, some models
 include other effects such as the metallicity dependence proposed by
 \cite{Kobayashi98}. The authors found that when the metallicity is
 lower than [Fe/H]$< -1$, the winds developed by the primary star (WD)
 accreting mass through the Roche lobe are too weak and the explosion
 cannot occur.

The second most popular scenario involving a thermonuclear explosion
of a C-O WD is the Double Degenerate (DD) scenario -- two WDs that
loose angular momentum and energy by emitting gravitational waves and
eventually merge. If the exceed the Chandrasekhar mass they ignite as
SNIa \citep{IbenTutukov84}.

Other suggested theoretical scenarios include the ``Collisional
Scenario'' where the head-on collisions of two WDs occurs instead of
the spiralling due to loss of gravitational wave radiation. These
collisions of two WDs of sufficiently large masses are predicted in
dense environments such as globular clusters, and could explain
supernovae occurring in the nuclei of galaxies
\citep{LorenIsernyelP10}. On the other hand, the ``Core-Degenerate
(CD) scenario'', where a WD merges with the core of an asymptotic
giant branch (AGB) star and forms a rapidly rotating WD. This new
configuration has a mass close to and above the critical mass for
explosion, and was used recently as the best scenario to explain the
observed properties of SN 2011fe \citep{Althaus13}. Additionally, the
``Double Detonation'' mechanism propose a sub-Chandrasekhar WD
accumulating a layer of helium-rich material on the surface
sufficiently massive and degenerate to cause a detonation
\citep{Shen13}. Finally, another type of Core-Degenerate scenario
considered is the ``super Chandrasekhar scenario''. These models
propose a Chandrasekhar or a super-Chandrasekhar mass WD formed in the
planetary nebula phase (or at the end of the common envelope phase),
from a merger of a WD companion with the hot core of a massive
asymptotic giant branch (AGB) star \citep{Tenebroso13}.  However, a
common characteristic of all these scenarios is that similar
nucleosynthesis processes occur during the explosion creating mainly
Fe. For a recent review on the SNIa progenitors see \cite{Maoz13}.

 Lately, many empirical DTDs have been proposed. The Bimodal model by
 \citet{Mannucci06} considers a DTD with two populations of
 progenitors of SNIa, one dominated by the ``prompt'' component that
 explodes within 100 Myrs after the formation of their progenitors,
 and the ``tardy'' component exploding on a wide period of time
 extending up to 10 Gyr .  The shape of the bimodal distribution is
 given by the sum of two functions: a Gaussian centered at
 $5\times10^{7}$yrs, and an exponentially declining plateau
 function. Moreover, similar power law DTDs ($\sim$t$^{-1}$) have been
 recovered from different observational surveys. For instance,
 \cite{Totani08} using field elliptical galaxies proposed such a
 power law. \cite{Maoz12} also suggested a very similar DTD based on a
 sample of 132 SNeIa from the Sloan Digital Sky Survey
 II. Finally, \cite{Pritchet08} presented a power law DTD
 ($\sim$t$^{-0.5}$) from the SuperNova Legacy Survey.

The DTD represents a powerful tool, not only for testing models and
helping to constraint the progenitor SNIa scenarios, but also for
obtaining an accurate description of the chemical and energetic
feedback from SNIa. 
The aim of our work is to test which is the best scenario for SNIa
in the framework of cosmological simulations.  In particular, we will
include and analyse in detail, for the first time to our knowledge,
five different DTDs for SNIa in a smooth particle hydrodynamical
code.  Our results will be presented in a series of three papers.

In this first paper we will discuss in detail the SD model
and compare our results, obtained for a bulge-type galaxy, with the
chemical abundances and SNIa rates for this kind of galaxy. In
particular, we will compare our results with the abundance data for
the Milky Way bulge.  We study global properties such as the
correlation found by \citet{Sullivan06} between the specific star
formation rate (SSFR) and the specific SNIa rate (SSNIaR).  The reason
for starting with this particular scenario is that it has been
suggested as one the best models to reproduce the chemical properties
of galaxies \citep{Matteucci06,MatteucciFelice09}.

\begin{figure}
\begin{center}
\includegraphics[width=0.55\textwidth]{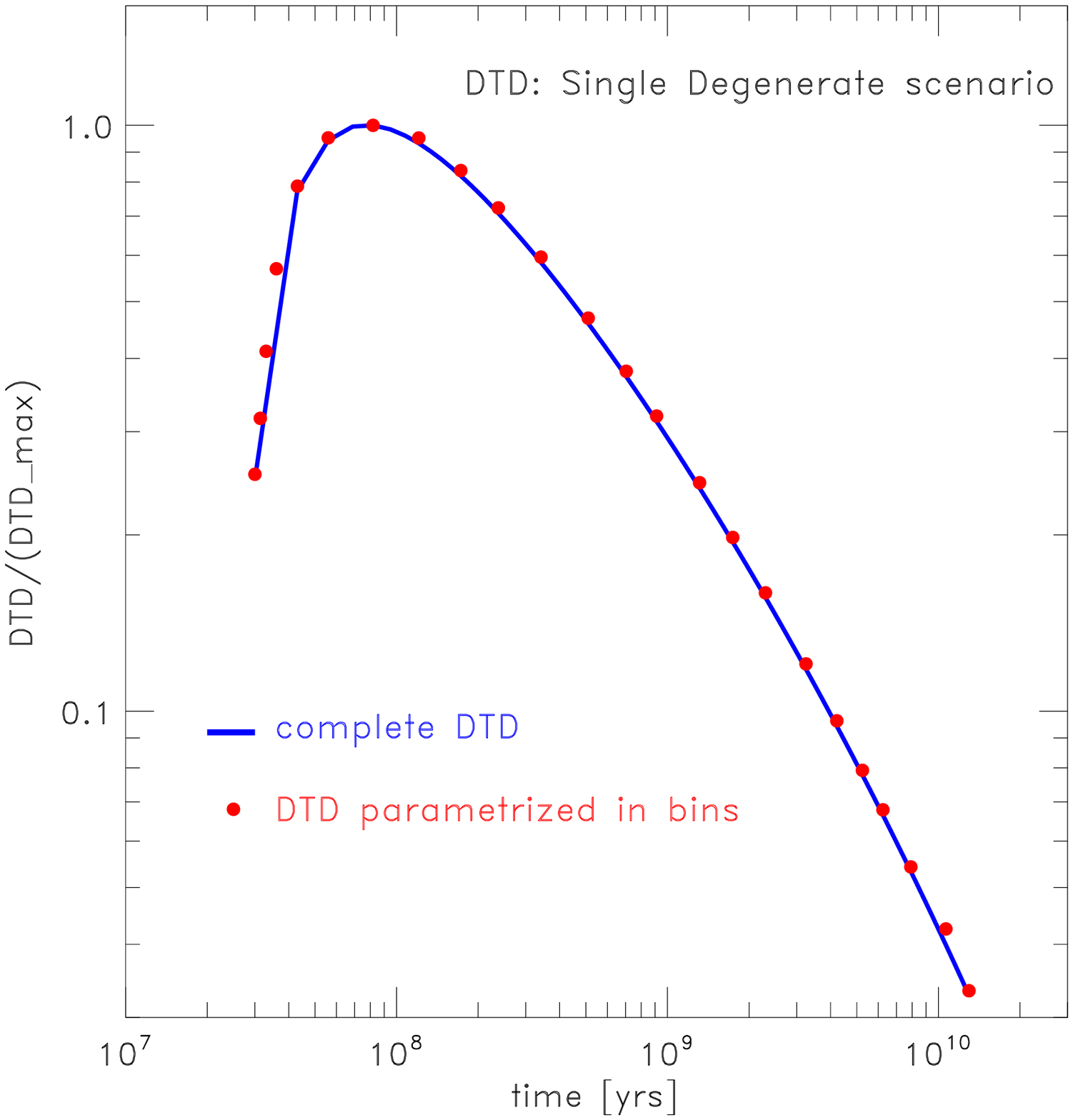}
\includegraphics[width=0.55\textwidth]{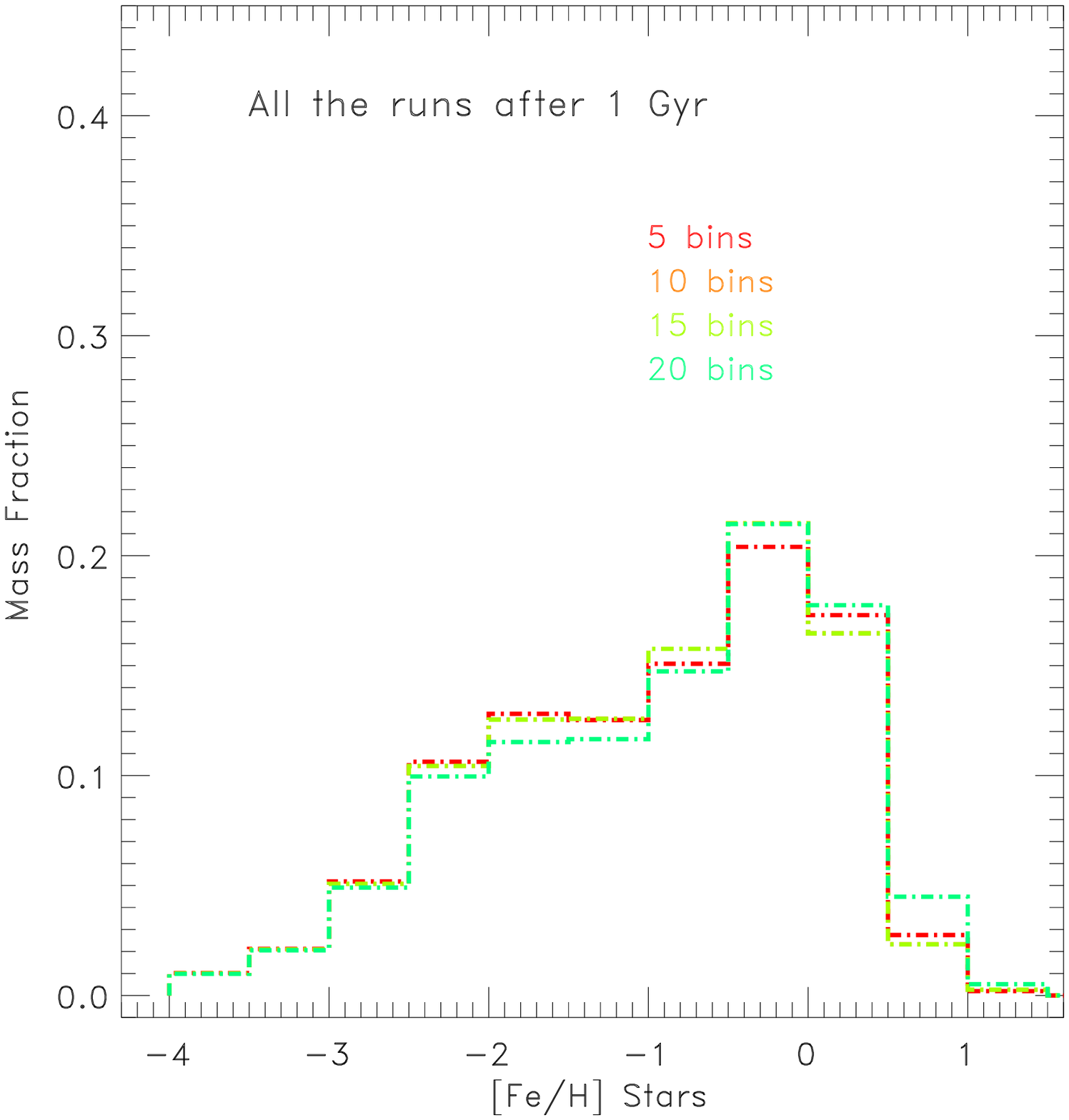}
\caption{ ({\it Upper panel}) The full DTD for the SD scenario given
  by \citet{MR01} and the DTD parametrized with 20 bins to reduce the
  computational costs (for more details, see the text). ({\it Lower
    panel}) Distribution of [Fe/H] ratios for the stars in the
  simulation, presented as an example of the relevant quantities of
  this study not being affected by the adopted number of bins used in
  the parametrization of the SD.}
\label{figConv}
\end{center}
\end{figure}

%In a second paper (Paper II), we will implement five DTDs (SD, DD,
%bimodal and two power laws) calibrated to reproduce the observables
%discussed in this paper. We will explore the \citet{Sullivan06}
%correlation for all the assumed DTDs.  In Paper III, we will show the
%differences in the metallicity between the stellar populations and the
%interstellar medium (ISM), when different DTDs are acting. 
% Our final
%goal is to run a cosmological volume.

 This paper is organized as follows: Section \ref{Sec:DTD} describes
 the analytic DTD for the SD scenario. Section \ref{Sec:code} briefly
 describes the main aspects of the numerical code and presents the
 initial conditions. In the following Subsections we present the
 implementation of DTD in the code. In Section \ref{Sec:calibre} we
 show the calibration of the model with the observables. In Section
 \ref{Sec:correlacion} we study the observed correlation found by
 \citet{Sullivan06}, between the specific SNIa rates and the specific
 SFR for the SD scenario.  The main conclusions are summarized in
 Section \ref{Sec:Conclusions}.

\section{The Single Degenerate scenario}\label{Sec:DTD}
 
\begin{figure}
\begin{center}
\includegraphics[width=0.45\textwidth]{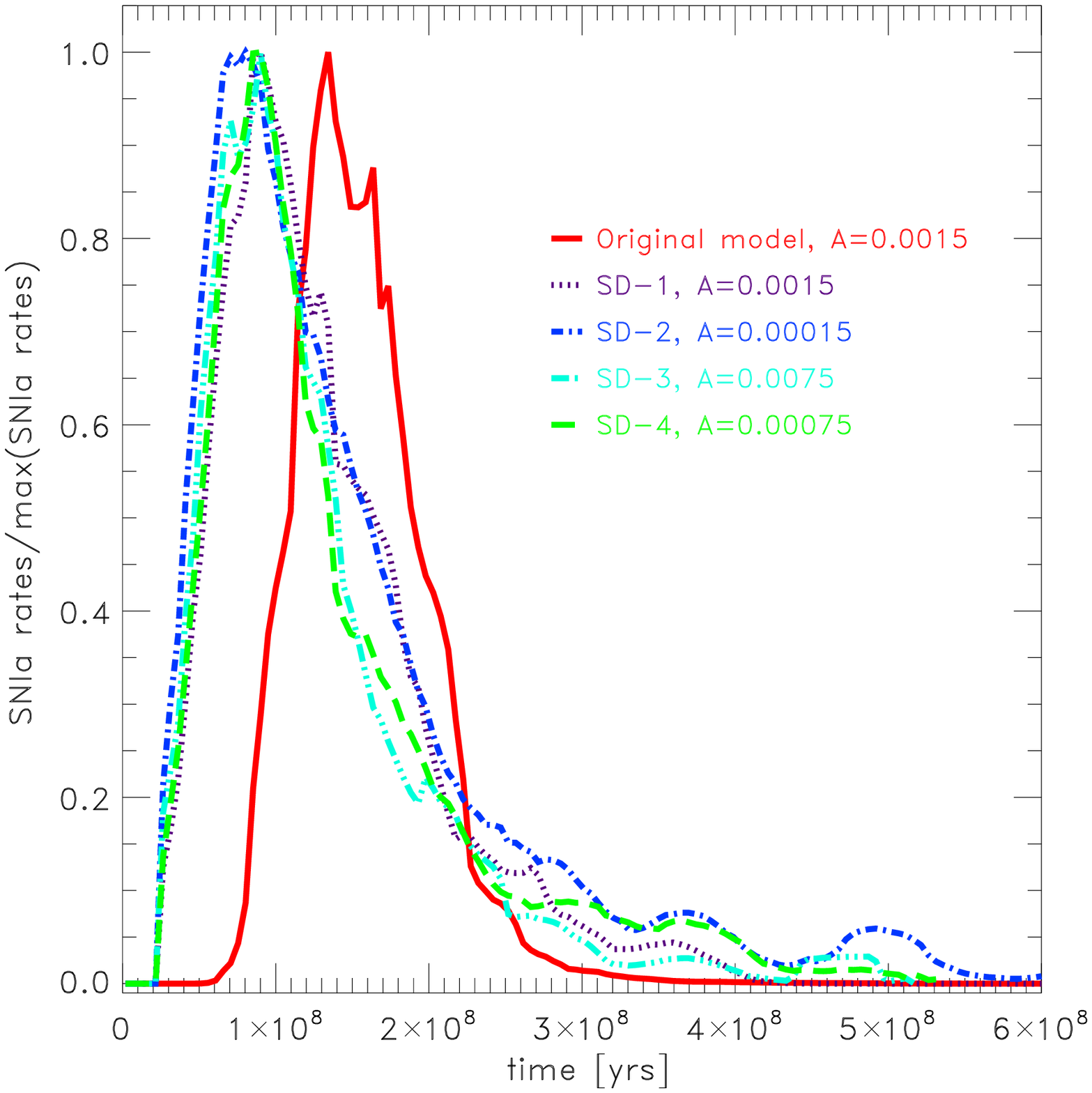}
\caption{SNIa rates for all the models as  function of time.}
\label{figSNIaR}
\end{center}
\end{figure}

In the SD proposed by \cite{WhelanIben73}, SNIa originate from a
binary system with one C-O WD and a red (super) giant star. The WD
accretes mass from the red giant through the Roche lobe, and an
explosive nucleosynthesis process occurs when the WD reaches the
Chandrasekhar mass. Calculations of the SNIa rates were performed on
the basis of this model
\citep{Greggio83,MatteucciGreggio86,Tornambe87,MatteucciFelice09} for
both our Galaxy and for elliptical galaxies.

To calculate the SNIa rates in the framework of the SD scenario is
necessary to know the range of lifetimes from the stars belonging to
the exploding binary systems. One way to do this is to evaluate the
minimum and maximum masses allowed for these systems, taking into
account the fact that Type Ia SNe occur in all galaxies, including
ellipticals which do not show any active star formation.  In this work
we adopt the model of \citet[from now GR83]{Greggio83}, where the
lifetimes of the stars exploding as SNIa are comprised in the mass
range of $0.8 - 8 M_{\odot}$. Then, according to this mass range,
after a starburst the first systems to explode as a SNIa
(e.g. 8+8$M_{\odot}$ systems) will require no more than $\sim
3-4\times10^{7}$ yrs. And this is the most massive system possible
with a total mass of $16 M_{\odot}$. On the other hand, the minimum
mass possible for a binary system giving rise to a SN Ia explosion is
estimated to be $3.0 M_{\odot}$ (see GR83). This total minimum mass
will ensure that the system can contain a white dwarf and a companion
massive enough to allow the white dwarf to reach the Chandrasekhar
mass after accretion.

\noindent For the  calculation of  the SNIa rates we follow the formulation of
\citet[from now, MR01]{MR01}:

\begin{equation}
R_{\rm Ia}(t)=A \int_{M_{\rm B, inf}}^{M_{\rm B, sup}}\phi(M_{\rm B})
\int_{\mu_{\rm min}}^{\mu_{\rm max}}f(\mu)
\psi(t-\tau_{M_2})d\mu\, dM_{\rm B}.
\label{eq2}
\end{equation}
\noindent here $M_B = M_1 + M_2$ is the total mass of the binary
system (where $M_1$ is the primary and $M_2$ the secondary).  In the
limits of the integral we use the minimum and maximum total masses for
the binary systems ($M_{\rm B, inf}$ and $M_{\rm B, sup}$) described
above. For more details see GR83 and MR01.

  The only free parameter of this formulation is A, representing the
  fraction of binary systems able to produce SNeIa in the mass range
  3-16$M_{\odot}$.  This parameter is generally fixed by reproducing
  the present time observed SNIa rate of the type of galaxy under
  study.

\noindent In Equation \ref{eq2}, the function $\psi(t)$ represents the
SFR. This quantity needs to be evaluated at the time $(t-\tau_{M_2})$,
where $\tau_{M_2}$ accounts for the life of the secondary star. This
constitutes the clock for the SNIa explosion
\citep{PadovaniMatteucci93}. The function $\phi(M_B)$ refers to the
Initial Mass Function (IMF) from \citet{Salpeter55}, defined in the
mass interval 0.1-100 $M_{\odot}$, with an index of $x=1.35$.

\begin{equation}
\phi(M_{\rm B}) = C M_B^{-(1+x)}
\end{equation}

\noindent
The mass fraction of the secondary star to the total mass of the
system is the quantity $\mu=M_2/M_{\rm B}$, and its distribution
function is given by $f(\mu)$. Previous studies \citep{Tutukov80}
suggest a value of $\gamma$=2, that we also adopt here, as previously
used in the formula:
\begin{equation}
f(\mu)=2^{1+\gamma}(1+\gamma)\mu^\gamma.
\end{equation}

\noindent It is worth noting that the integral in Equation \ref{eq2} when 
computed without the SFR, namely for an instantaneous starburst, is
the so-called Delay Time Distribution (DTD) function (shown for the SD
in top panel of Figure \ref{figConv}).

\noindent This approach allows us to evaluate the Equation \ref{eq2},
for each stellar particle with a SF episode and estimate the rate of
SNIa.

\section{The numerical code}\label{Sec:code}
\begin{sloppypar}

We use an extended version of the Tree-PM SPH code GADGET-3
\citep{Springel05}, which includes metal-dependent cooling
\citep{ShuterlandDopita93}, star formation, chemical enrichment,
supernova feedback (SN), and a multiphase model for the gas component
\citep{SC06}. In this scheme if two gas particles have dissimilar
thermodynamic properties, they are explicitly prevented from being
neighbours in the SPH calculations (unless they are in a shock),
allowing the coexistence of gas clouds with different thermodynamical
properties. The SN feedback scheme is coupled adequately with the
multiphase model. As a result, the energy injection into the ISM
produces the self-regulation of the SF and the triggering of
mass-loaded galactic outflows. This SN feedback does not introduce
mass-dependent parameters as described in detail by \citet{SC06}.
In this work, we assume the total SN energy release in each event to
be $0.7\times 10^{51}$ erg and is equally distributed within the cold
and hot phases surrounding a stellar particle.  Our SN feedback model
has already been used to study the formation of disc galaxies in a
cosmological context \citep[e.g.][]{Tissera13, TisseraP13b,Artale14}.
\end{sloppypar}

The cold and dense gas is transformed into stars when satisfying
density and temperature criteria according to the Kennicutt-Schmidt
law \citep{KSlaw0, KSlaw1}. We adopt a Salpeter IMF, defined in the
mass interval of 0.1-100\Msun. SNe type II are assumed to originate
from stars more massive than M$>8$\Msun and lifetimes of $\sim 10^6$~
yrs.
\begin{sloppypar}
  The SNIa prescription included in \citet{SC06} was originally
  proposed by \citet{Mosconi01}. Regarding the SNIa implementations,
  the model of \citet{SC06} will be referred as the ``original model''
  to distinguish it from the new implementations for the SNIa (based
  on different DTDs), presented in this and the following papers.  In
  the original model the lifetimes of the binary system that explode
  as SNIa are assumed to be randomly distributed within a certain
  range given by $\tau_{SNIa}=[0.1,1]$ Gyr \footnote{These limits can
    be varied as has been done by \citet{SC09} in the Aquarius
    simulations, for example, thus considering a larger minimum
    lifetime for the distribution of the secondary mass, compared to
    the predictions of the SD scenario. This constitutes a simple
    approach based on the assumption that a fair fraction of SNIa will
    explode during a period $\tau_{SNIa}=10^{8}-10^{9}$ yrs
    \citep{Greggio96}.  A fixed relative ratio between SNII and SNIa
    is assumed adopting an observationally motivated value
    \citep{vandenBergh91}, which can be associated with the free
    parameter A.  The chemical yields in all the models are give by
    the W7 model of \citet{Iwamoto99}}.  This SNIa scenario, albeit
  simple, has been successful at reproducing many observational
  chemical patterns and trends as shown in \cite{Tissera12,Tissera13,
    Tissera14}.  The original model has the large advantage of being
  computationally inexpensive while grafting the main features of
  SNIa.  However, there is room for improving the SNIa modeling as
  explained in the Introduction. This would be of a great importance
  when large on-going or planned surveys of the our Galaxy or nearby
  galaxies start to yield detailed abundances of the stellar
  populations such as APOGEE (a part of the Sloan Digital Sky Survey
  III) \citep{Anders13}, The GIRAFFE Inner Bulge Survey (GIBS)
  \citep{Zoccali14}, The Calar Alto Legacy Integral Field Area
  (CALIFA) survey \citep{Sanchez13}, The Sydney-AAO Multi-object
  Integral field spectrograph (SAMI) \citep{Croom12}, among others.
\end{sloppypar}

\subsection{The simulated galaxies}\label{Subsec:IC}

\begin{figure}
\begin{center}
\includegraphics[width=0.55\textwidth]{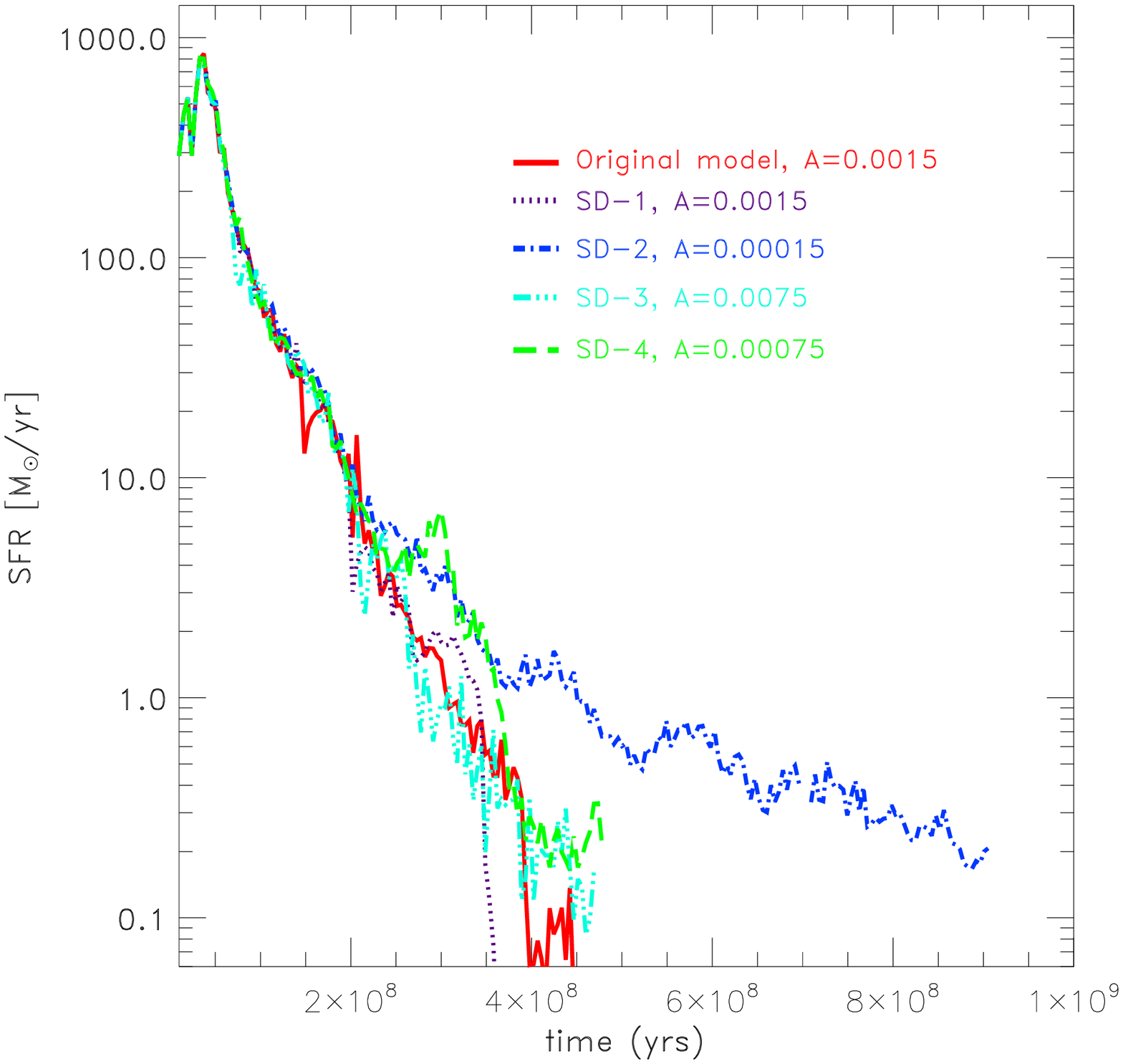}
\includegraphics[width=0.55\textwidth]{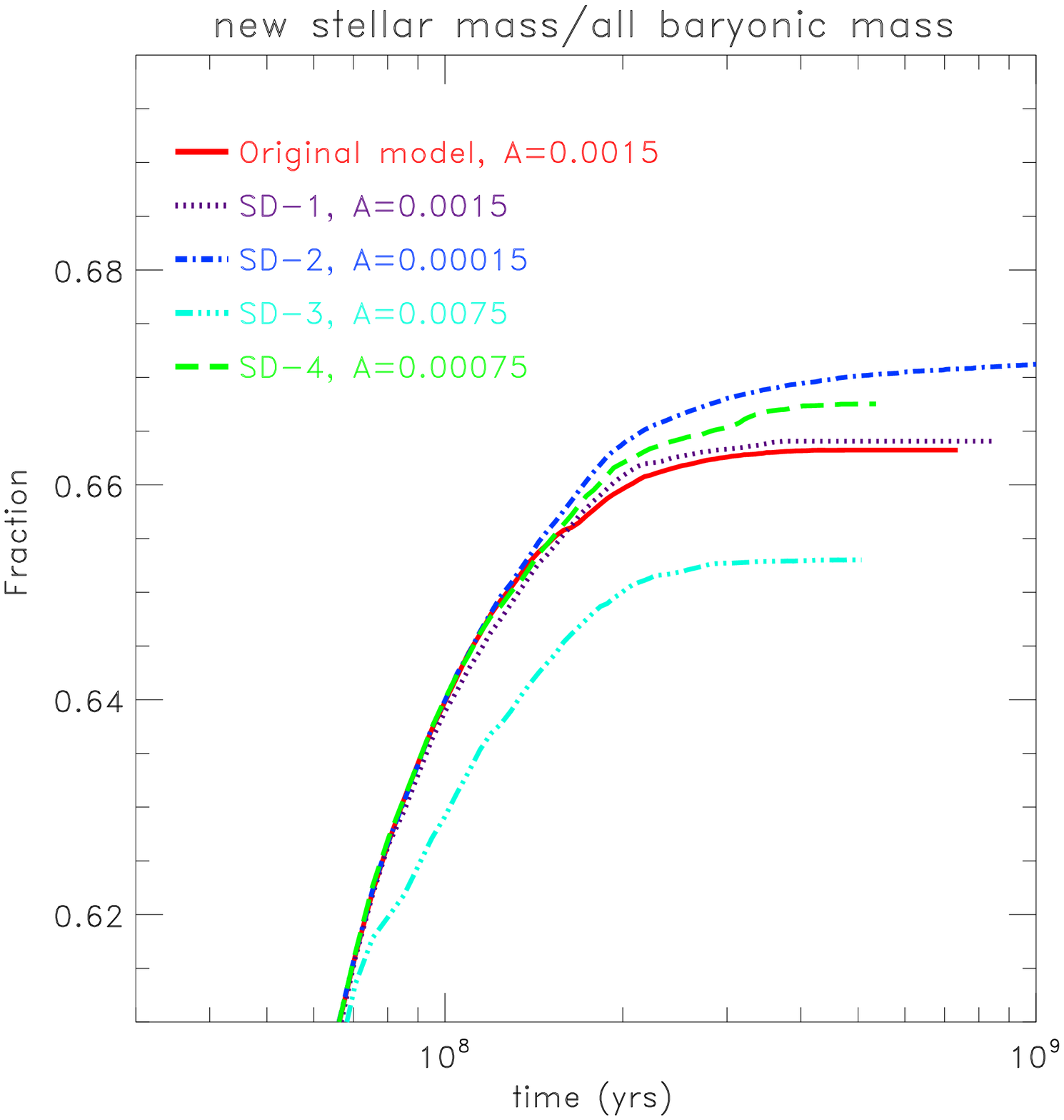}
\caption{SFR history (upper panel) and mass fraction of new born stars
  (lower panel) for the SPH galaxies run with the same IC
  and varying the A values.}
\label{figSFR}
\end{center}
\end{figure}

We analyse the performance of the DTDs in simulations of pre-prepared
galaxies in isolated dark matter haloes.  This initial condition (IC)
is simple enough to highlight the effects of the different DTDs
without being distracted by additional processes such as mergers and
gas infall, which complicate the picture in fully hierarchical
scenarios for galaxy formation. This simple approach allows us to more
easily test the influence of the free parameter of the implemented
DTDs.

The IC consists of a dark matter potential with an initial
distribution following a NFW density profile \citep{NFW97}, which is
let to evolve self-consistently with the baryonic component, with a
concentration of $c=9$, an old stellar bulge with a Hernquist profile
and a old stellar exponential disc. The virial mass of this system is
M$_{200} \sim 10\times10^{11}$\Msun, with $10\%$ of this mass in form
of baryons.  The adimensional spin parameter is $\lambda=0.044$.

Initially, the gas is distributed in the disc component and represents
$\sim 65\%$ of the total baryonic mass of the galaxy. Within the total
disc component, the gas represents a fraction of $\sim 90 \%$ of the
initial mass of the disc. This initial large gas fraction has been
chosen to mimic a galaxy in its first stages of evolution in a simple
way.  At the beginning of the simulations the disc component is in
equilibrium within the potential well (Q $> 1$) and has a scale-length
of $\sim 3.4 $ kpc. The gaseous disc has a lower Q value (closer to
$\sim 1$) and becomes unstable developing clumps and bar
instabilities. These two perturbations transport material into the
inner regions of the galaxy contributing to the formation of the bulge
component. In particular, the bar drives gas inflows which trigger
important starbursts \citep[for more details
  see][]{Perez11,Perez13}. 

 In all the simulations the initial number of baryonic particles is $
 \sim 80000$ and the dark matter particles $\sim
 100000$. Specifically, the initial gas particle is $\sim
 7\times10^{5}$\Msun~ while the dark matter particle is
 $9\times10^{6}$\Msun~ and the stellar mass particle, $\sim
 3\times10^{5}$\Msun. The softening length for the gas particles is
 $200$ pc and for the dark matter particles we adopt $450$ pc.  The
 SFR efficiency is set at $c=0.1$. We follow the evolution of the
 systems until the gas to form stars is depleted and the star
 formation ceases.

 We decomposed the bulge and disc components of the simulated galaxies
 by adopting a dynamical criterion based on the ratio between the
 rotational and dispersion velocities \citep[e.g.][]{SC08}. From this
 analysis we identify a bulge and disc component. The bulge is formed
 by a dissipative-dominated component and a central bar. The stellar
 density profiles show that the bulge is concentrated within the inner
 $\sim 3$ kpc. This configuration shows a ratio of $B/T \sim
 0.84$. Hence, the new stars formed for which we follow the chemical
 abundances are mainly part of the bulge.

Even more, the analysis of the SFR of the bulge shows that the stars
formed during the strong starbursts, where the cold gas is exhausted
within $\sim 1$ Gyr.  The new stars remain concentrated within $\sim
3$ kpc. All the simulations show this strong initial SF burst ending
before $\sim 1$ Gyr due to the gas depletion into stars, and the
effects of SN feedback which ejects part of the gas (see Subsection
\ref{Subsec:SFR}). The B+D stellar mass is nearly the same for all the
simulations $\sim 3.5\times10^{10}$\Msun, typical of a small
ellipticals or spiral galaxy bulges.

  By comparing these SF histories with the ones given by the model of
  \citet[][hereafter PM04]{PM04}, we conclude that the SF history of
  the gas component resembles spheroidal-type systems. This is
  consistent with the fact that the gas becomes unstable and collapses
  into the central region, feeding the strong starbursts. Most of the
  new stars are centrally concentrated, contributing to the formation
  of the bulge component.  Therefore, we compare the present day SNIa
  rates with those observed in spheroidal-dominated galaxies. The
  chemical patterns will be confronted with observations of the Galactic
  Bulge as these are the only observations available of individual
  stars in a bulge. We do not attempt to reproduce exactly these data
  since the  formation history of the Galactic Bulge seems to be
  more complicated \citep{RojasA14}, but to use the data to set global
  constrains on the A parameter.

\begin{table}
  \caption{Main characteristics of the simulated galaxies run with the
    SD scenario and varying A. The mean observed SNIa rate for
    spheroidal dominated galaxies of stellar mass $\sim
    3.5\times10^{10}$\Msun is $\sim 0.0017$ SN per year \citep{Li11}.}
  \begin{center}
    \begin{tabular}{l|l c c  c}
        Model & A &  $<$SFR$>$ & SNIa rates \\\hline
        & & [\Msun yr$^{-1}$] & [N/yr$^{-1}$] \\\hline
       SD-1      &0.0015   & 83 & 0.002\\ 
       SD-2      &0.00015  & 60 & 0.00006\\
       SD-3      &0.0075    & 84 & 0.0080 \\
       SD-4      &0.00075   & 76 &  0.0016\\
       original  &0.0015  & 53 & 0.0027\\ \hline
     \end{tabular}
  \end{center}
  \label{table1}
\end{table}

\subsection{Implementation of the DTD}\label{Subsec:implemente}

The numerical code estimates whether if a given gas particle meets the
condition to be transformed into stars. Then, the code follows over
time the new stellar particle, representing a single stellar
population, and calculates the number of SNIa which should be produced
as a function of time, according to the assumed DTD. The free
parameter A is adjusted to reproduce the present-time observed SNIa
rates. Analytical or semi-analytical models fix the value of A
requiring the galaxy to reproduce the observed SNIa rates according to
its morphology. In a numerical simulation, A is tuned at a particle
level, so a given stellar particle does not have any a {\it priori}
knowledge of the morphology of the galaxy it inhabits. On the
contrary, reproducing the observed SNIa rate according to morphology
should be a prediction of our models. Note that in our models, gas and
star particles will evolve according to the physical laws in a
non-linear way. Hence, reproducing the observed values is a challenge,
even with a IC as simple as the one adopted in this work.

As mentioned in the Introduction, there are several theoretical
scenarios for the DTDs. This paper considers the SD by MR01 and leave
for Paper II the following scenarios: DD by \citet{Greggio05}, the
Bimodal scenario by \citet{Mannucci06}, the power laws proposed by
\citet{Maoz12}, and by \citet{Pritchet08}. The implementation of the
DTDs is the same for all the models.

The drawback of including detailed descriptions of the DTDs in the
simulations is the computational cost for our particular SN feedback
model. The objective is indeed to keep the model running efficiently
to allow its use in large-scale cosmological simulations.  Therefore,
solving the integrals given by Equation \ref{eq2} for each particle,
at each time, is highly inefficient. This is related to the
methodology followed by SN feedback model. In the adopted scheme, the
code has to search for the nearest hot and cold gas particle
neighbours of a given stellar particle in order to inject the energy
and chemical elements. This procedure is very time-consuming. To
alleviate this problem, we include the DTDs in the simulations by
creating tables for each DTD. We use 20 equally time-spaced bins to
parametrize the complete DTD distributions. This can be visualized in
the upper panel of Figure \ref{figConv}, where we show the complete
DTD for the SD scenario given by MR01 (continuous line), and the
re-sampling with 20 intervals or bins (red dots). The representation
of the DTD with a this lower number of intervals reduces the
computational costs by limiting the hot and cold neighbours searchings
required to inject the SNIa products, without losing the relevant
information stored in the complete DTD distribution.  The SNIa rates
for each stellar particle (i.e. representing a single stellar
population) are estimated by searching in the DTD table for the number
of SNIa which should be produced according to the particle stellar age
and the adopted progenitor model. Then, the total SNIa rates for the
central spheroid are calculated by adding the contributions form each
star particle identified to belong to this component. In particular,
we use the rates at the time when the SFR is quenched to compare with
other models and observations. Thanks to this efficient implementation
based on pre-prepared tables for the DTDs, the computational costs are
significantly reduced.

\begin{sloppypar}
We study the convergence of different parametrizations of the DTDs
varying the bins numbers used to map the DTDs.  We run the SD-1 model
with $ [5, 10, 15, 20]$ number of bins for fitting the DTD curve,
finding that the distribution of metals, the SFR, specific SFR, the
SNIa rates, and other relevant quantities for this study do not not
change significantly with the number of bins chosen. To illustrate
this fact we show in the right panel of Figure~\ref{figConv},
histograms of the mass fraction of stars within a given [Fe/H],
calculated for models with different numbers of bins. It can be seen
that the differences between the fractions of mass are
negligible. Notice, for example, that the fraction of stars with
[Fe/H]$>0$ vary less than $0.2\%$ between the models. 
\end{sloppypar}

\begin{figure}
\begin{center}
\includegraphics[width=0.55\textwidth]{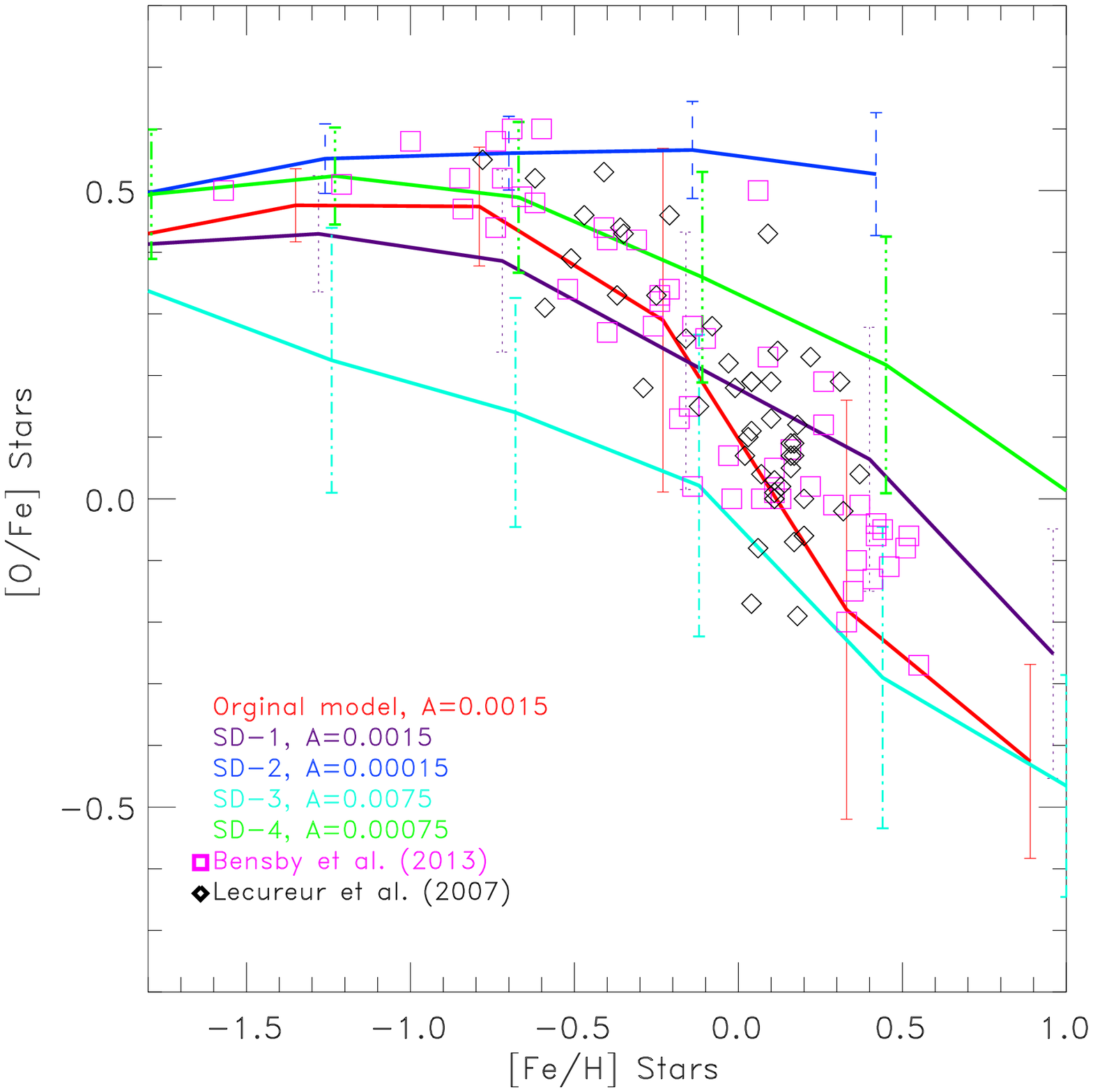}
\caption{[O/Fe] vs. [Fe/H] exhibited by the bulge stars in the SPH
  simulated galaxies with the SD scenario by \citet{MR01} and varying 
  A. These models are compared to observational [O/Fe] ratios for
  stars in the Galactic Bulge by \citet{Bensby13,Lecu07}. See Table
  ~\ref{table1} for details on the simulations.}
\label{figZocalo}
\end{center}
\end{figure}
\placefigure{figZocalo}

\section{Calibrating the SD scenario}\label{Sec:calibre}

Chemical evolution models calibrate the DTDs using the free parameter
A fixed a {\it posteriori}, to fit the present day observed rate of
SNIa.  This parameter accounts for the fraction of binary systems that
undergo a SNIa event in a starburst once the IMF is fixed. In the SPH
simulations A is also a free parameter but on particle basis. Thus is
not possible to know {\it a priori} in which morphological type of
galaxy a given particle inhabits, particularly in a cosmological
simulation.  The parameter A has to reflect the underlying physics
and, at the same time, be able to reproduce the mean observable
constraints.

To explore the range of A values able to reproduce observations, we
make use of the multi-zone chemical evolution model of PM04 for
bulge-like systems of similar mass and SF history as the SPH
galaxies. Once we find a parameter space for A, we run the SPH
simulations.  The advantage of this methodology is the small
computational cost of running the multi-zone chemical evolution model
of PM04, compared to the cost of running several SPH simulations until
selecting the proper values for A. Caution should be taken since a
change in A might produce a non-linear response in the SPH
simulations, as the SN feedback and gas metallicities will also
change, modifying the cooling rates and the availability of cold gas
for subsequent star formation activity.

Specifically, in the PM04 model a galaxy is divided into several
non-interacting shells. The chemical evolution equations are solved in
each shell in order to reproduce the evolution of the elemental
abundances. The SFR is given by the law $\psi(t )=\nu \rho_{gas}(t )$,
assumed to be proportional to the gas density $\rho_{gas}$, via a
constant $\nu$, representing the star formation efficiency. This
quantity in the PM04 models increases with the baryonic galaxy mass.
For the comparison with the SPH simulations, we run the model of PM04
adopting $\nu=50$ Gyr$^{-1}$ for a galaxy with initial mass of
$10^{11}$\Msun and a infall time of $0.01$ Gyr. Because of the stellar
winds suffered through its evolution, the galaxy ends up with a final
mass of $\sim 4.7\times10^{10}$\Msun. And this mass is of the order of
magnitude of the new stellar mass formed in the SPH galaxies ($\sim
3.5\times10^{10}$\Msun).  The SFR history of the SPH galaxies and the
PM04 galaxies are very similar, showing a bursty behaviour at the
beginning with a fast quench after $\sim 1$ Gyr.  These SFRs are shown
in Figure~\ref{figSFR}, upper panel.

 The following step is to use the SFR of the SPH galaxy obtained
 running the original model of \citet{SC06} as an input to the
 model of PM04. In this way, we obtain a prediction for the rates of
 SNIa and SNII for the given galaxy. By varying the value of A in the
 PM04 model, we can find the adequate parameter space for A.  This
 parameter space is then the one we adopt in the SPH simulations. As a
 constraint to our models we use the rate-size relation of \cite{Li11}
 and derive the observed SNIa rate at the present time, following the
 method shown by \cite{Valiante09}. We do this for a galaxy with
 baryonic mass similar to our simulated galaxy. Thus, for a system of
 $\sim 3.5\times10^{10}$\Msun, the rate predicted by \cite{Li11} is
 $\sim 0.0017$ SNIa per year.
\begin{sloppypar}
 We run four experiments with the same IC and the SD model using
 A $=[0.0015, 0.00015, 0.0075, 0.00075]$ (SD-1, SD-2, SD-3 and SD-4
 models). Larger or smaller values are strongly ruled out by the PM04
 model.  Table \ref{table1} summarizes their main properties, including
  the predicted SNIa rates (column 4), calculated when the SFR is
  quenched.  It can be seen that we obtain good agreement with SD-1
  and SD-4.  The original model, with $A=0.0015$, also reproduces the
  observed SNIa rates.
 \end{sloppypar}
 The behaviour of the SNIa rates as a function of time for the
 different models is shown in Figure~\ref{figSNIaR}. It becomes clear
 from the plot that not only the A parameter affects the rates, but
 also the lifetimes of the progenitor systems that explode as SNIa
 given by each model. Recall that for the SD scenario the first
 systems made of two $8 M_{\odot}$ stars explodes after $\sim
 3-4\times10^{7}$ yrs after the starburst. On the other hand, for the
 original model the lifetimes are assumed to be randomly distributed
 within a certain range given by $\tau_{SNIa}=[0.1,1]$ Gyr. Thus, the
 predicted rates for the original model are delayed in time relative
 to those of the SD, and lacking of the prompt SNIa (those exploding
 in the first 0.1 Gyrs), as can be seen in Figure~\ref{figSNIaR}. In
 the following section we analyze the consequences of this delay in
 the SFR.

\subsection{Star Formation Rate}\label{Subsec:SFR}

In Figure ~\ref{figSFR} (upper panel), we show the SFR of the
simulated galaxy for all the explored A values. The SFRs are estimated
by using constant time intervals along the history of the galaxies. In
all the runs, there is strong burst at the beginning, followed by a
decay of the SFR until the gas is exhausted.  Since the simulated
galaxy is isolated, there is no external gas reservoir to keep feeding
the SF.  No meaningful differences between the models appear until the
SFR declines below $\sim 10$\Msun yr$^{-1}$. The SFR for the SD-1
model is the first to drop to $\sim 0.01$\Msun yr$^{-1}$ after $\sim
4\times10^{8}$ yrs. The model SD-2, with the lowest A value, is the
last to exhaust the gas by forming stars after $\sim 8\times10^{8}$
yrs. Models SD-1 and the original share the same value of A but show
different quenching times. The latter being the one with more extended
SFR. All the simulations have their SFR quenched before 1 Gyr, due to
gas depletion into stars and/or the effects of SN feedback, which
heats up part of it.

Alternatively, we quantify the effects of the parameter A acting on
the regulation of the SF by the estimation of the mass fraction of
new-born stars. Figure~\ref{figSFR} (lower panel) shows clearly how
different A can regulate the SF activity according to the total energy
that SNeIa inject into the ISM.  The efficiency of the stellar
production depends on the available cold gas in galaxies, a quantity
linked to the number of SN events occurring in the galaxy. From
Figure~\ref{figSFR} we can see that when the value of A increases the
fraction of new stars decreases accordingly.  This can be clearly
appreciated from the comparison between the final fractions reached in
SD-3 and SD-2, with the highest and lowest values of A,
respectively. However, the variations between models are very small
(less than $2\%$) indicating that, at least for this IC, the effects
on the regulation of the SF are minor, if sensible values of A are
assumed.

Comparing the original and SD-1 models we can see that they produce
the same fraction of new-born stars, although the star formation
activity is quenched earlier in the SD-1 model. Recall that in the SD
model the minimum lifetime of SNIa is $\sim 3\times10^{7}$ yrs,
whereas in the original model is $\sim 1\times10^{8}$ yrs (see Figure
~\ref{figSNIaR}). This explains the differences in the SFR quenching
times of the two models. Notice that the DTD implementation seems not
to affect the efficiency of the fraction of new-born stars formed in
the galaxies when compared to the original model. Observational
evidence of the existence of prompt SNeIa strongly argues in favour of
a minimum time-scale as long as $\sim 10^{8}$ yrs \citep[and
  references therein]{Bonaparte13}.

 Notice that even when the SNIa can continue to pollute the ISM with
 the nucleosynthesis production after the SFR has stopped, this will
 not affect the chemical abundances of the stellar populations since
 no new stars will be formed from the enriched gas.  Therefore, in the
 following Section we analyse the distribution of chemical elements at
 this particular time: when the SFR is quenched.  Analytical and
 multi-zone chemical evolution models adopt a similar criterion that
 we also chose here for the sake of comparison.

\subsection{The [O/Fe] vs. [Fe/H] diagram }\label{Subsec:AphasSD}

 We study here the predicted [O/Fe] vs. [Fe/H] relation by each of the
 models of the SD scenario included in the SPH simulations. It is
 worth remembering that the parameter A is tuned to reproduce the
 present time observed SNIa rate. In our case, the values of A which
 best reproduce the present time rate are those of SD-1 and SD-4
 models, as already shown in Section \ref{Sec:calibre}. Here we test
 the effect of varying A, namely the fraction of SNIa, on the
 predicted [O/Fe] ratios.\footnote{In the original and SD models the
   yields used for SNeII are those metal-dependent from
   \citet{WW95}. For the production of SNIa, we assume the W7 model by
   \citet{Iwamoto99}.}

We interpret the evolution of the abundance ratios of [O/Fe] vs [Fe/H]
with the help of the time-delay model of \citet{Tinsley79}. It is
expected that the delay in the production of Fe by SNIa ejected into
the ISM -- in relation to the rapid production of $\alpha$ elements by
SNII -- leaves a characteristic signature in the [O/Fe] vs [Fe/H]
diagram MR01. The ejection of Fe from SNIa is regulated by
the DTD. The main effect of the delayed Fe production in relation to
the $\alpha$-elements produced by SNII, is to create an
over-abundance of $\alpha$-elements. In particular, the Oxygen remains
high until SNIa becomes important and the ratio [O/Fe] starts to
decline. This point is identified by a ``knee'' in [O/Fe] vs [Fe/H]
diagram (see Figure \ref{figZocalo}).

Assuming that bulges and elliptical galaxies experience a strong burst
of star formation lasting a short time (less than 1 Gyr)
\citet{MatteucciBrocato90} predicted that [$\alpha$/Fe] should be
super-solar for a large interval of [Fe/H]. In fact, objects such as
bulges that evolve very fast with an intense SFH, quickly reach solar
metallicity only due the production of Fe by SNeII. When SNeIa start
to explode, the production of Fe is enhanced and a change in the slope
(the knee) in the [O/Fe] vs [Fe/H] diagram occurs at [Fe/H] $\ge 0$.

 We compute the average stellar mass abundance of [Fe/H] and [O/Fe]
 for the stars in the bulge (see Sec. \ref{Subsec:IC}), for each of
 the models. The ratios [Fe/H] and [O/Fe] are calculated by adding the
 masses of the corresponding chemical elements stored in the star
 particles identified to belong to the central spheroid.  We show the
 SD models with different values of A in Figure~\ref{figZocalo}, where
 the different coloured lines refer to each of the models compared to
 Galactic Bulge stars. A sample of dwarfs and subgiant stars from
 \citet{Bensby13}, and Red Giants stars from \citet{Lecu07}, are
 considered for comparison.

  Figure~\ref{figZocalo} shows the observed data points lying in
  between the curves displayed by models SD-3 and SD-2, corresponding
  to the highest and lowest values of A, respectively. However, model
  SD-2 predicts a too flat [O/Fe] ratio. This means that the fraction
  of SNIa is too low. On the other hand, model SD-3 is an extreme case
  where there are too many SNeIa and it predicts, in fact, a
  continuous decrease of the [O/Fe] ratio, at odds with the
  observations.  The data are best represented by the models SD-4 and
  SD-1. The former model (SD-4) fits the zero point of the data
  following the observed trend up to [Fe/H]$\sim -0.25$. Meanwhile,
  SD-1 matches the slope and passes through the data better. These
  models predict a long plateau for the [O/Fe] ratio and a knee
  occurring at high [Fe/H], as observations suggest. However, as can
  be seen in the Figure ~\ref{figZocalo}, at high metallicity the
  slopes of both models do not so nicely follow the observed [O/Fe]
  ratios. This is perhaps a consequence of yields adopted in this
  work. The WW95 yields do not include mass loss from massive stars,
  which is particularly important for Wolf-Rayet stars and for
  supersolar metallicity. Its effect is to increase the yields of
  Carbon and Helium and to depress that of Oxygen, as extensively
  described in \cite{McWilliam08}.  Notice that SD-1 and SD-4 models
  also fit the present time SNIa rates (see Section
  \ref{Sec:calibre}).  The original model, albeit simple, fits the
  data in the whole range.  

From the previous analysis, we conclude that the chemical enrichment
shown by the [O/Fe] ratios changes linearly with A in the SD
scenario. Therefore, it is possible to find a range of values of A
which can predict the current SNIa rates and the expected trend for
[O/Fe] ratios at the same time.  It also shows the importance of
calibrating the models using observables. 

\begin{sloppypar}
The [O/Fe] vs [Fe/H] ratios observed in the solar neighbourhood had
been reproduced for Milky Way-type (MW) galaxies by several galaxy
formation models \citep[e.g.][]{Calura12,Yates13,Few14,DeLucia14},
although some of them fail to reproduce the most enriched starts for
the MW bulge, showing an offset with the data \citep[see for example
  Figure 12 from][]{DeLucia14}.

The ``knee'' of the bulge shown in Figure~\ref{figZocalo} occurs
around [Fe/H]$\sim -0.25$, at variance from the observed value for the
disc component of the Milky Way found at [Fe/H]$\sim -1$
\citep{Francois04}.  This difference is a consequence of the strong SF
experienced by the spheroidal component, making it evolve faster than
the disc. As a result, it reaches higher values of [Fe/H] at the time
when the SNeIa stars to become important and restore the bulk of Fe to
the ISM \citep[a prediction from chemical evolution
  models][]{Matteucci90}. Therefore, we recall the reader than the
time-scale for the SNIa reaching its maximum enrichment, usually
quoted as $\sim$ 1 Gyr, is not universal but only valid for the solar
neighbourhood, as can be seen in this example and was pointed out by
MR01.
\end{sloppypar}

\section{The correlation between the SNIa rates and the specific SFR}\label{Sec:correlacion}

 Observations relating the SNIa rates of galaxies with the
 characteristics of the host galaxy such as their morphology, colours
 and SFR are powerful constraints to our models. Consequently, in this
 Section we compare the results of the best SD scenario with
 observations presented by \citet{Sullivan06}.  In that paper, the
 authors found a correlation between the Specific Star Formation Rate
 (SSFR) -- the SFR per unit mass of the galaxy --, and the specific
 SNIa (SSNIa) -- the rate of SNIa per unit of galaxy mass -- for
 galaxies in the SuperNova Legacy Survey (SNLS) galaxy sample. This
 correlation is based on a sample of 100 spectroscopically confirmed
 SNeIa, plus 24 photometrically classified events, distributed over
 $0.2 <$ z $< 0.75$. The stellar masses and SFRs for the SNIa host
 galaxies are estimated by fitting their broadband spectral energy
 distributions (SEDs) with the galaxy spectral synthesis code PEGASE.2
 \citep{Pegase97}. They adopt the IMF of \citet{Kroupa01}.  The SSFR
 for the sample is measured as the ratio between the mean SF rate over
 the last 0.5 Gyr and the current stellar mass of the galaxy,
 resulting from the SED fitting. They choose this interval of time to
 avoid systematic errors for galaxies for which the redshift is not
 known.  

Furthermore, \citet{Sullivan06} compared the correlation with
observations of a morphologically classified sample of SNeIa host
galaxies in the local universe presented by \citet{Mannucci05}. It
became clear from this comparison that the sample of galaxies in the
local universe shares the same trend displayed by SNLS galaxies at
higher redshift. Therefore, the SSNIaR is a function of the host
galaxy SSFR, with strongly star-forming galaxies hosting roughly 10
times as many SNeIa per unit mass than passive galaxies with no star
formation.

 Moreover, \citet{Smith12} obtained the SNIa rates of galaxies with
 different SFR activity located at intermediate redshift ($0.05 <$ z
 $< 0.25$), from a sample of 342 galaxies belonging to the Sloan
 Digital Sky SuperNova Survey-II (SDSS-II SN Survey). The authors
 estimated the host stellar masses and the recent star-formation rates
 using the code PEGASE.2. They confirmed the existence of a
 correlation between the SSFR and the SSNIaR. Therefore, the
 correlation holds for intermediate redshifts \citep{Sullivan06}, and
 for the local Universe \citep{Mannucci05}, indicating no evolution
 with redshift within the range of $0.05<z<0.75$.

\citet{Sullivan06} concluded that the correlation is difficult to
reconcile with a model for SNIa that originates solely from an old
evolved stellar population. Instead, they proposed a scenario of two
separate components: a prompt component with a short delay time and an
old component with a long delay time, consistent with the Bimodal
model of \citet{Mannucci06}.  Here we explore this relation with the
SD model and in Paper II the analysis will be extended to the other
DTD models.

To assess if our best SD models are able to reproduce a correlation
such as the observed one, the SSNIaR and the SSFR are estimated as the
galaxy evolves, under the hypothesis that observations might catch
galaxies at different stages of evolution (see also Figure 12 in
\cite{Greggio10} for a similar approach using analytical models). For
this purpose, we calculate the SSNIaR and the SSFR of the simulated
galaxies as a function of time. This is done by using the SFR
histories and the SNIa rates of the simulated galaxies and their
corresponding stellar masses as function of time (e.g. Figure
~\ref{figSFR}).  These quantities allow us to calculate the mean SSFR
and SSNIaR at different stages of the galaxy evolution. The averages
over the whole stellar population are taken at a certain period of
time, chosen to be within 0.5 Gyr, in order to mimic
observations. \footnote{ To correct for the different IMFs used in
  this work (Salpeter) and the observations (Kroupa), we adopt the
  transformation suggested by \cite{Longhetti09}.}
\citeauthor{Smith12} noticed that the zero point of the correlation is
affected by the choice of the IMF, systematic uncertainties concerning
the accuracy of the derived properties of host galaxies and the
photometric redshift estimates produced by PEGASE SED fitting
code. Therefore, because of all these uncertainties, we focus on the
study of the slope of this correlation and prefer not use the error
bars provided by the observational works.  Consequently, the simulated
relations are normalized by an {\it ad hoc} factor to simplify the
qualitative comparison of the observed slope of the correlation.

\begin{figure}
\begin{center}
\includegraphics[width=0.55\textwidth]{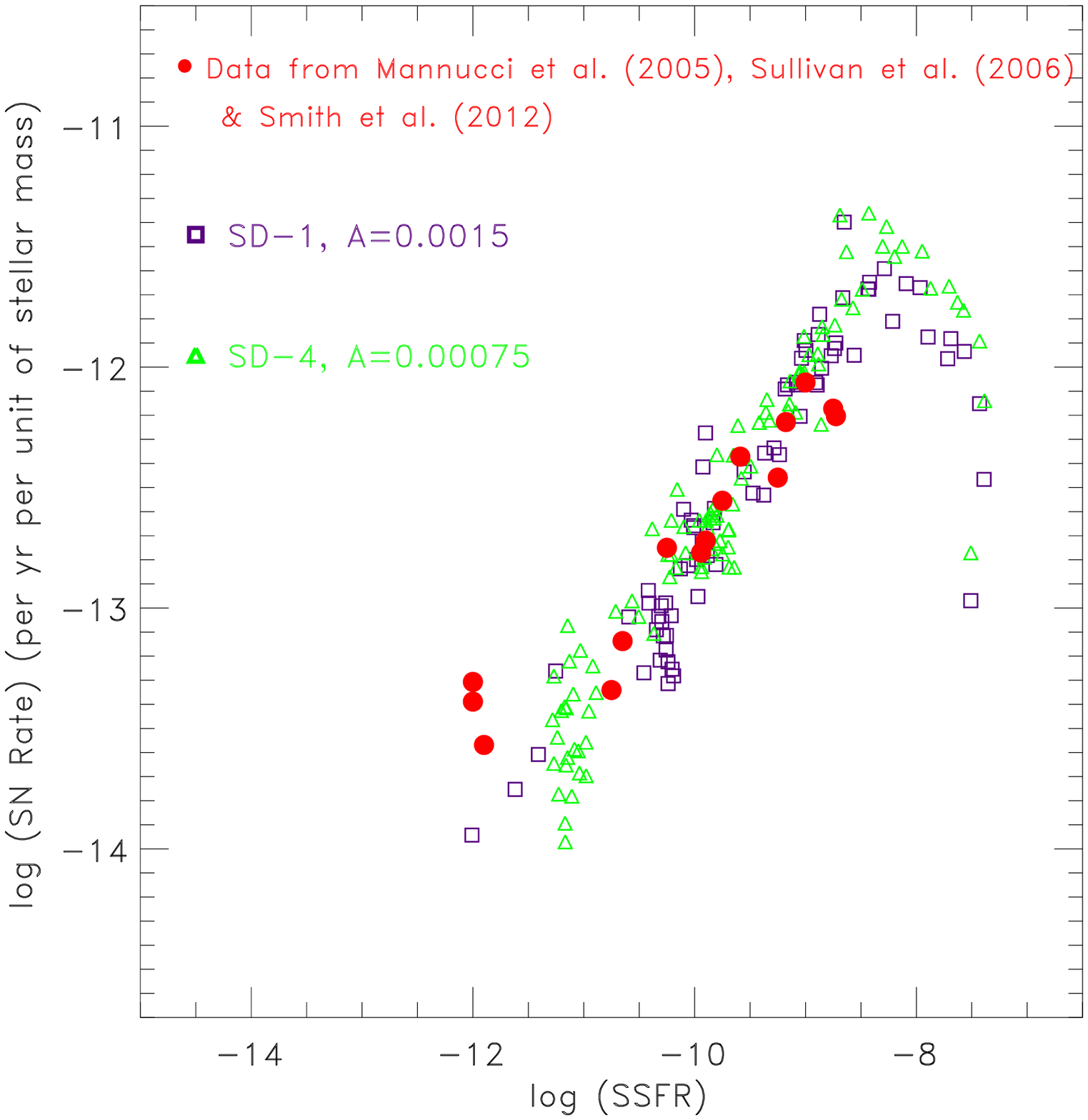}
\caption{Specific star formation rate (SSFR) as a function of the SNIa
  rate per unit of galaxy mass (SSNIaR) for the SD-1 and SD-4 models
  (runs with A$=0.0015$ and A$=0.00075$). The red circles represent
  combined data from \citet{Sullivan06,Mannucci05,Smith12}. Notice
  that these models also fit the observed [O/Fe] ratios coming from
  the Galactic Bulge and the present-day SNIa rates. The zero-point
  has been renormalized by an {\it ad hoc} factor. See Section
  \ref{Sec:correlacion} for further details.}
\label{figSullivan}
\end{center}
\end{figure}

\begin{figure}
\begin{center}
\includegraphics[width=0.55\textwidth]{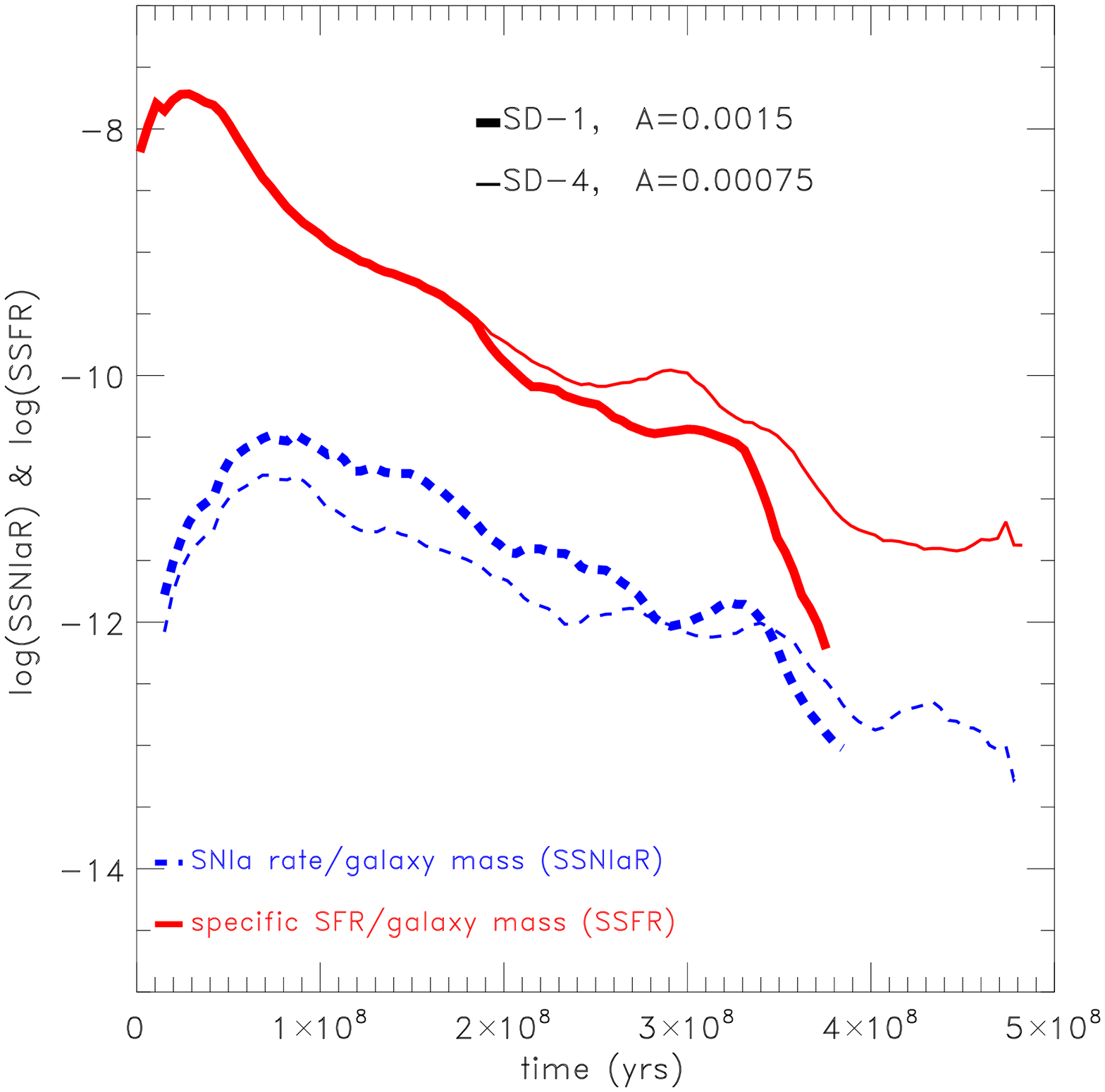}
\caption{The log(SSFR)--continuous line--and log(SSNIaR) --dashed
  line--, evolving with time for our best models within the SD
  scenario (SD-1 and SD-4).}
\label{figBoth}
\end{center}
\end{figure}

\begin{sloppypar}

A combination of the observed data by
\citet{Smith12,Sullivan06,Mannucci06}, and the results for the SD-1
and SD-4 models are shown in Figure~\ref{figSullivan}.  Although a
word of caution is necessary when making comparison between the
evolution in time of single objects -- like our simulated galaxies --
and observations referring to a mix of different objects that might
have undergone different evolutionary histories, it is interesting
that our model including the SD scenario reproduces a clear
correlation which agrees remarkably well with observations.
\end{sloppypar}
 As can be seen from Figure~\ref{figSullivan}, at the beginning the
 simulated correlations follow a different trend, so that the SSNIaR
 increases abruptly to high values of SSFR before reaching the
 expected observed trend. This feature is detected for all A values in
 the SD model, as can be seen in models SD-1 and SD-4.  These strong
 variations in the SSNIaR are produced in a short interval of SSFR for
 most of the models ($8.5<$log SSFR $<7.5$) and are related to the
 time-delay in the SNIa explosions.

To understand the physical reasons causing the SSFR-SSNIaR correlation,
we plotted the logarithm of SSFR and the SSNIaR as a function of time
for the SD-1 and SD-4 models in Figure~\ref{figBoth}.  The sharp
increase of SSNIaR for high SSFR reflects the onset of SNIa and how
quickly it reaches the maximum value. After that, both the SSNIaR and
the SSFR decrease establishing the observed correlation.  For very low
SSFR, there is still a residual SSNIaR, as the generation of these
events is delayed in time and the correlation is lost.

  The particular feature seen at the beginning of the SSFR-SSNIaR
  relation is not shown by the current observational data and
  represents a prediction from the simulations. If these early stages
  could be observed, then they could be used to set limits on the
  shape of the DTDs. Of course, the system we have used to explore the
  effects of varying the DTDs is very simple. The SF history of
  galaxies could certainly be more complicated with several starbursts
  occurring during their life. The analysis of these complex stellar
  populations is delayed to a forthcoming paper using cosmological
  simulations.

\section{Discussion and Conclusions}\label{Sec:Conclusions}

We present results from simulated galaxies performed with a version of
{\small Tree-PM SPH-GADGET-3} \citep{Springel05, SC06}, to study the
impact of the SNIa feedback by using different DTDs. In this paper, we
explore the SD model of MR01 in detail.  We choose the SD scenario as
a reference to discuss the implementation, the main observables and
correlations which we aim to reproduce.  The implementation of the SD
scenario involves the calibration of a free parameter A which
represents the fraction of binary systems in one stellar generation
giving rise to SNIa events. Chemical evolution models generally fix A
according to the SFR and the present day SNIa rates of the galaxy
under study. However in a cosmological simulations A acts at a
particle basis. Thus, it is assumed to be the same for all single
stellar populations, while the final SNIa rates or any relation
between the SNIa rates and the galaxy properties -- or the chemical
abundance patterns -- should come out as a prediction of the
simulations.  Therefore, it is necessary to explore a range of
suitable A values and their impact on the properties of galaxies, by
using simple initial conditions. In a forthcoming paper, we explore
other progenitor scenarios. The final goal is to run cosmological simulation and
analyse the chemical properties of galaxies of different masses and
assembly histories.  Adequately reproducing the SNIa feedback as well
as their effect on the chemical enrichment is very important, as new
Galactic surveys will start yielding high precision measurements of
chemical abundances.

Our main results can be summarized as follows:
 
\begin{itemize}
\item 
We find that the SFR responds linearly to the number of SNeIa in the
SD scenario. The SFR is found to decline faster with increasing values
of A. But the differences are small and so the final stellar mass
formed is very similar.  There is a substantial difference between the
original model of \citet{SC06} and the SD of \citet{MR01}
implemented and analysed here. This is related to the adopted
lifetimes for the secondary masses. In the original model, the SNIa
production was shifted to later times, compared to the SD scenario,
beginning at $\sim1\times10^{8}$yrs. This excludes the so-called
prompt SNeIa, that are indeed observed. For a similar A, the original
model shows a more extended SFR. However, both of them predict similar
final fractions of new stellar mass to total baryonic mass.\\

\item The SFR in the simulations is dominated by a strong starburst
  consuming the cold gas within $\sim1$ Gyr and forming stars that end
  up concentrated within the bulge.  Hence, the SNIa rates are
  comparable to observed ones for spheroidal galaxies. This comparison
  with observed SNIa rates of \cite{Li11} shows that the best fits are
  obtained for SD-1 and SD-4 models, adopting A$\sim 0.0015$ and
  A$\sim 0.00075$, respectively.\\

\item The [O/Fe] ratios predicted for the stars in the bulge of the
  simulated galaxies are compared to data from the Galactic Bulge
  (although this comparison is just indicative). We find that the best
  agreement with observations is again provided by the SD-1 and SD-4
  models. These models predict a long plateau for the [O/Fe] ratio and
  a knee occurring at high [Fe/H], as observations suggest. We note
  that this is the first time that such an excellent agreement is
  found in galaxy simulations.\\

\item 
At variance with previous claims, we find that the SD scenario (SD-1
and SD-4 models) reproduce the observed correlation between the SSNIaR
and the SSFR, found by \citet{Sullivan06}, if we estimate these
quantities at different evolutionary times. This correlation comes out
naturally from the simulations. Moreover, two features in the
correlations are shown which cannot be confronted with current
observations but could be interesting to explore in the future:

\begin{itemize}
\item 1) For high SSFR, as the SSNIaR starts to appear, the
  SSNIaR-SSFR anti-correlates before it turns into the observed
  correlation.  This is caused by the large initial number of SNe Ia
  in the assumed DTD. The turnover occurs when the maximum in the DTD
  is reached and hence, it could be interesting to explore if this
  turn-over could be confirmed with observations coming from galaxies
  dominated by a very recent starburst.

\item 2) For very low SSFR, there is still a residual SSNIaR because
  the majority of these events are delayed in time, thus the
  correlation disappears. The SSFR at which this occurs could be
  related to the shape of the DTD.

\end{itemize}
\end{itemize}

In a forthcoming paper, we will discuss other DTDs implemented in the
SPH simulations: the DD scenario by \citet{Greggio05} and empirically
motivated DTDs such as the Bimodal scenario by \citet{Mannucci06}, the
power laws reported by \citet{Maoz12} and \citet{Pritchet08}.  We will
explore the predicted SNIa rates and chemical abundances, as well as
the global correlations which can contribute to the understanding of
SNeIa progenitors.

\section*{Acknowledgments}\label{sec:gracias}
 NJ is very grateful for fruitful discussions and comments from
 Marcelo Miller Bertolami, Ilaria Bonaparte, Emanuele Spitoni, Laura
 Greggio, Jordi Isern, Nikos Pranzos, Robert Yates, Manuela Zoccali,
 Thomas Bensby, Enrique Gaztaniaga, Emanuel Sillero and Kai
 Hoffmann. Many thanks to Kate Rowlands, Jack A. Kennedy for their
 comments and corrections to the manuscript. We thank the anonymous
 referee for a very constructive report that helped to clarify the
 paper.

We thank Mark Sullivan and Mathew Smith  for kindly providing the data
shown in this paper.

 NJ acknowledges support from CONICET-Argentina; the European Research
 Council Starting Grant (SEDmorph; P.I. V. Wild); the European
 Commission's Framework Programme 7, through the Marie Curie
 International Research Staff Exchange Scheme LACEGAL
 (PIRSES-GA-2010-269264) and warmly thanks the Observatory of Trieste
 and the Institute of Space Studies of Catalunya, where most of this work was done. 

This work was partially supported by PIP 2009/0305(CONICET-Argentina)
and PICT Raices 2011/959 (Foncyt-Argentina). Simulations where run in
Fenix Cluster of Institute of Astronomy and Space Physics
(IAFE-Argentina) and Hal Cluster of the Universidad Nacional of
Cordoba (Argentina).  Also, this work was  supported by PIP
2009/0305(CONICET-Argentina), PICT Raices 2011/959 (Foncyt-Argentina),
``Proyecto Interno'' of the Universidad Andres Bello (Chile), and
Fondecyt Regular 2015 1150334.  F.M. acknowledges financial support
from PRINMIUR2010-2011, project: ``Chemical and dynamical evolution of
the Milky Way and Local Group Galaxies'', prot.N. 2010LY5N2T. This
research made use of the NASA's Astrophysical Data Systems and
arXiv.org e-print service from the Cornell University.

\bibliography{aamnem99,Manuscript3}

\end{document}